\documentclass[12pt]{article}
\usepackage{natbib,graphicx,amsmath,amsthm,amssymb}
\usepackage{comment}

\renewenvironment{itemize}
{\begin{list}{$\bullet$}{\setlength{\topsep}{0cm} \addtolength{\leftmargin}{-10mm} \setlength{\itemsep}{0cm}\addtolength{\labelwidth}{-10mm} }} {\end{list}\medskip}

\renewenvironment{description}{\begin{list}{}{\setlength{\topsep}{0pt}  \setlength{\itemsep}{0pt}
 \labelwidth 0pt \itemindent -32pt \addtolength{\leftmargin}{-5mm}  }} {\end{list}\medskip}

\newcommand{\eg}{{e.g.}\ }

\usepackage{color}

\begin{document}

\vspace*{2cm} 

{\Large\textbf{Global simulation envelopes for diagnostic plots in regression models}\\

\begin{minipage}{\textwidth}
\large
{David I.\ Warton}

\end{minipage}

\normalsize

School of Mathematics and Statistics and Evolution \& Ecology Research Centre, UNSW Sydney, NSW 2052, Australia

ORCID: 0000-0001-9441-6645




\newpage

\begin{quote}
Residual plots are often used to interrogate regression model assumptions, but interpreting them requires an understanding of how much sampling variation to expect when assumptions are satisfied. In this paper, we propose constructing global envelopes around data (or around trends fitted to data) on residual plots, exploiting recent advances that enable construction of global envelopes around functions by simulation. While the proposed tools are primarily intended as a graphical aid, they can be interpreted as formal tests of model assumptions, which enables the study of their properties via simulation experiments. We considered three model scenarios --  fitting a linear model, generalized linear model or generalized linear mixed model -- and explored the power of global simulation envelope tests constructed around data on quantile-quantile plots, or around trend lines on residual vs fits plots or scale-location plots. Global envelope tests compared favorably to commonly used tests of assumptions at detecting violations of distributional and linearity assumptions. Freely available \texttt{R} software (\texttt{ecostats::plotenvelope}) enables application of these tools to any fitted model that has methods for the \texttt{simulate}, \texttt{residuals} and \texttt{predict} functions.

\end{quote}

\section{Introduction}
When fitting a regression model, residual plots are typically relied upon as a diagnostic tool to check assumptions are reasonable. However, a challenge for interpretation is that residuals will always stray from expected trends to some extent -- we would like to know if deviations from expected patterns are large relative to sampling variation.

One way to compare observed patterns to expected sampling variation is to construct a global envelope -- a region of the diagnostic plot within which we expect the observed data to always lie if assumptions are satisfied (at a chosen level of confidence $1-\alpha$). \citet{Rosenkrantz2000} developed an approach to construct global confidence bands around a quantile-quantile (or ``QQ"-) plot, under the assumption that observations are independent and normally distributed, by exploiting Mood joint confidence sets for the mean and variance. \citet{Aldor-Noiman2013} devised a closely related method, but used simulation to estimate the desired multiple testing correction needed for global envelopes.

Many regression models do not assume errors are independent and normally distributed, as for example when analysing discrete data (Figure~\ref{fig:egPlots}). In such cases simulation could be used to construct envelopes, repeatedly simulating new responses from the fitted model, refitting the model and constructing residuals. \citet{Moral2017} used this approach to generate pointwise simulation envelopes around half normal plots. Note however that pointwise envelopes can be quite misleading, as they do not correct for multiple testing. The chance that not all observations fall inside a pointwise envelope can be much higher than the nominal rate, so na\"ive interpretation of pointwise envelopes could too often lead to the conclusion that assumptions have been violated, in situations where they have not been violated. What is needed here is to construct global simulation envelopes -- controlling the chance that \emph{all} observations are contained inside the simulation envelope, across datasets that satisfy model assumptions.

This paper describes a method of constructing global simulation envelopes for residual plots from any fitted model, using the method of \citet{Myllymaki2017}. Their method was designed as a diagnostic tool for point patterns, but in principle, it can be used to construct global simulation envelopes around any function. Their method is used in this paper to construct global simulation envelopes around three commonly used types of residual plot (Figure~\ref{fig:egPlots}), as a visual tool to aid interpretation of residual plots. Such envelopes may also be useful as a teaching tool, to assist students in learning when departures from expected trends are large relative to sampling variation. A global envelope could also be used in a formal test of assumptions, and while this is not the primary intention of the proposed tool, it does enable study of its properties via power simulation. Simulations presented here suggest that global envelopes on residual plots have power that is competitive with formal tests designed to interrogate model assumptions. The proposed global envelope approach has been implemented on \texttt{R} as the \texttt{plotenvelope} function in the \texttt{ecostats} package \citep{WartonBook}, and can in principle be applied to \emph{any} model object that has methods for the \texttt{simulate}, \texttt{residuals} and \texttt{predict} functions.

\section{Global envelopes}

Assume we have a diagnostic plot which defines a function $T(r)$ over some set of values $r\in \mathcal{I}$, which could be an interval, or a discrete set of values. In constructing a $100(1-\alpha)\%$ level global envelope over $\mathcal{I}$, we want to find functions $T_\text{low}(r)$ and $T_\text{upp}(r)$ such that when the assumed regression model is satisfied
\[ \mathbb{P}\left[ \exists r\in \mathcal{I}: T(r)\notin (T_\text{low}(r), T_\text{upp}(r)  )\right] = \alpha \]

\subsection{Constructing global envelopes via simulation}
\citet{Myllymaki2017} proposed a method of constructing a global envelope for $T(r)$ via simulation -- making inferences from a set of simulated realisations of the function $T_b(r)$ for $b=2,\ldots,B$, where we additionally include $T_1(r)=T(r)$, the observed function, in this functional sample. While \citet{Myllymaki2017} were motivated by the problem of diagnosing goodness-of-fit in point process models, \eg using Ripley's \citeyearpar{Ripley1976} $K$-function, the approach could equally well be applied to any type of function. Here we will define functions using common residual diagnostic plots, and simulate new residuals by simulating data from the fitted regression model, refitting the regression model, and recomputing the residuals \citep[a ``parametric bootstrap",][]{davison1997bootstrap}.

The core idea \citep{Myllymaki2017} is to construct a global statistic that measures the maximal distance $u$ of each function $T_b(r)$ from the centre of the $B$ functions $T_0(r)$, and construct envelopes such that the $(1-\alpha)$-quantile of $u$ is only exceeded with probability $\alpha$. For example, we could compute the maximum absolute difference (MAD) measure, as compared to the mean function, and construct global envelopes based on absolute distance from $T_0(b)$. Specifically, if we let the mean function be $T_0(r)=\frac{1}{B}\sum_{b=1}^B T_i(r)$, then for each function $T_b(r)$ we compute $u_b=\max\limits_{r\in \mathcal{I}} |T_b(r)-T_0(r)|$. We can then construct a $100(1-\alpha)$\% global envelope as
\[\left(T_0(r)-u_{1-\alpha},T_0(r)+u_{1-\alpha}\right)\]
where $u_{1-\alpha}$ is the $(1-\alpha)$-quantile of the $u_b$, computed as the $(1-\alpha)B$-smallest value of $u_b$. Since $u_{1-\alpha}$ is only exceeded by a proportion $\alpha$ of the $u_b$, only a proportion $\alpha$ of the simulated functions $T_b(r)$ will stray outside this envelope.

An issue with using MAD is that $T(r)$ may have non-constant variance -- in fact for most diagnostic plots we consider below, we would expect $T(r)$ to have a smaller variance for intermediate values of $r$. We will compute the pointwise sample variance function $\text{var}(T(r))=\frac{1}{B-1}\sum_{b=1}^B \left( T_b(r)-T_0(r) \right)^2$ and measure maximal standardized distance between $T_b(r)$ and $T_0(r)$, computed as $v_b=\max\limits_{r\in \mathcal{I}} \frac{|T_b(r)-T_0(r)|}{\sqrt{\text{var}(T(r))}}$. Hence we construct a Studentized MAD global envelope \citep{Myllymaki2017}:
\[ \left(T_0(r)-v_{1-\alpha}\sqrt{\text{var}(T(r))} ,T_0(r)+v_{1-\alpha}\sqrt{\text{var}(T(r))} \right) \]
where as before $v_{1-\alpha}$ is the $(1-\alpha)$-quantile of the $v_b$. Up to Monte Carlo error, this envelope provides an exact global test of a simple hypothesis, where the distribution of simulated data has been exactly specified \citep{Myllymaki2017}. It has been noticed empirically that the method is often conservative for composite hypotheses, where the distribution we wish to sample from is a function of unknown parameters \citep{Myllymaki2017}. When using global simulation envelopes to diagnose regression models, we find ourselves in this latter situation. Software to compute Studentized MAD global envelopes is available on \texttt{R} in {\verb+GET::global_envelope_test+} using the option \texttt{type="st"}.

\subsection{Diagnostic plots in regression models}

Now consider the situation where we have univariate responses stored in a vector $\mathbf{y}$ of length $n$, with the $i$th observation denoted $y_i$ for $i=1,\ldots,n$. 
We can define vectors of linear predictors $\hat{\boldsymbol{\eta}}=g(\hat{\boldsymbol{\mu}})$ and residuals $\boldsymbol{e}$, whose $i$th elements are $\hat{\eta_i}$ and $e_i$. The residual $e_i$ is some pre-defined function of $y_i$ and $\hat{\eta_i}$, an ideal choice being a function such that the distribution of residuals does not vary as a function of linear predictors when assumptions are satisfied. There are many different types of diagnostic plot that could be constructed. Below we describe three commonly used plots, and how we define the interval $\mathcal{I}$ and the function $T(r)$ for each plot:
\begin{description}
\item[QQ-plot:] This is a plot of sorted residuals against theoretical normal quantiles. Let $e_{(i)}$ be the $i$th residual when placed in rank order. In a normal quantile or ``QQ"-plot we plot $e_{(i)}$ against $z_i=\Phi^{-1}\left(\frac{i-0.5}{n}\right)$ for $i=1,\ldots,n$, where $\Phi(z)=\mathbb{P}(Z\leq z)$ is the cumulative distribution function of the standard normal distribution $Z\sim\mathcal{N}(0,1)$. Inverting this formula for $z_i$, we get $i=n\Phi(z_i)+0.5$.

    We define as our function the sorted residual, as a function of its corresponding normal quantile. Specifically, we let $\mathcal{I}$ be the set of normal quantiles $\{z_1,\ldots,z_n\}$, and $T(r)=e_{(n\Phi(r)+0.5)}$.
\item[Residuals vs fits plot:] We plot residuals $e_i$ against linear predictors $\hat{\eta}_i$, for $i=1,\ldots,n$. If there is no violation of the mean model, residuals $e_i$ should have no trend, as a function of linear predictors. We can diagnose this by fitting a smoother to the residuals, as a function of linear predictors:
    \[ e_i = f(\hat{\eta}_i)+\epsilon_i \]
    where $\epsilon_i$ are errors centred on zero, and $f(\cdot)$ is estimated using some non-parametric or semi-parametric regression method. We will estimate $f(\cdot)$ using thin plate regression splines \citep{Wood2003}, assuming the $\epsilon_i$ are Gaussian with constant variance, using \texttt{mgcv::gam} on \texttt{R} with maximum likelihood estimation \citep[\texttt{method="ml"}]{Wood2011}.

The function we will use as a diagnostic tool for residual vs fits plots is the smoother defined above, $T(r) = f(r)$, acting on the interval containing the observed range of linear predictors, $\mathcal{I}=\left( \text{min}(\hat{\eta}_i), \text{max}(\hat{\eta}_i) \right)$.

\item[Scale vs location plot:] To look for violation of variance assumptions it is common to plot the absolute value of residuals $|e_i|$ against linear predictors $\hat{\eta}_i$, for $i=1,\ldots,n$. 
    If there is no violation of the mean model, there should be no trend. We can diagnose this by fitting a smoother to $|e_i|$, as a function of linear predictors:
    \[ |e_i| = f_s(\hat{\eta}_i)+\epsilon_i \]
    where as previously $\epsilon_i$ are errors centred on zero, and we will estimate $f_s(\cdot)$ using \texttt{mgcv::gam(\ldots, method="ml")}.

The function we will use as a diagnostic tool for scale-location plots is the smoother defined above, $T(r) = f_s(r)$, acting on the interval containing the observed range of linear predictors, $\mathcal{I}=\left( \text{min}(\hat{\eta}_i), \text{max}(\hat{\eta}_i) \right)$.
\end{description}

\subsection{Global envelopes for residual plots}
We have described how to construct a global envelope via simulation, using the methods of \citet{Myllymaki2017}, given a set of simulated functions $T_b(r)$, $b=1,\ldots B$, and we have defined three different types of function $T(r)$ based on different residual diagnostic plots that could be used to query goodness-of-fit of a regression model $\mathcal{M}(\widehat{\boldsymbol{\theta}})$ that has been fitted to sample data $\mathbf{y}$. We will construct simulated realisations of these functions by simulating data $\mathbf{y}_b$ from the fitted regression model and recomputing the relevant residual function $T_b(r)$, for each $b=1,\ldots,B$, as follows:
\begin{itemize}
\item Simulate data $\mathbf{y}_b$ from fitted regression model, $\mathcal{M}(\widehat{\boldsymbol{\theta}})$
\item Refit the regression model to simulated data, to obtain $\mathcal{M}(\widehat{\boldsymbol{\theta}}_b)$
\item Compute residuals $\mathbf{e}_b$ for the simulated data $\mathbf{y}_b$ from the refitted model $\mathcal{M}(\widehat{\boldsymbol{\theta}}_b)$.
\item Hence compute $T_b(r)$ from resampled residuals $\mathbf{e}_b$ and (if required) the observed linear predictors $\widehat{\boldsymbol{\eta}}$.
\end{itemize}
This procedure is sometimes referred to as a parametric bootstrap \citep{davison1997bootstrap}. Note that this procedure requires us to fit the model $\mathcal{M}(\boldsymbol{\theta})$ to data $B$ times, which can potentially be quite computationally intensive, if the model cannot be fitted quickly.

In residual vs fits and scale-location plots, note that linear predictors were not resampled, but were kept fixed at their observed value $\widehat{\boldsymbol{\eta}}$ as computed using the regression model fitted to the observed data $\mathcal{M}(\widehat{\boldsymbol{\theta}})$. Initial simulations considered resampling the linear predictor -- computing $\widehat{\boldsymbol{\eta}}_b$ from $\mathcal{M}(\widehat{\boldsymbol{\theta}}_b)$ to use in residual plots -- but in noisy data settings, the variability in $\widehat{\boldsymbol{\eta}}_b$ often suppressed the signal in subsequent residual plots.

\subsection{Software implementation in the \texttt{ecostats} package}
Software to implement this approach is available from the Comprehensive R Archive Network \citep{CRAN} in the \texttt{ecostats} package, which was written to accompany an introductory text on regression modelling \citep{WartonBook}. The key function is \texttt{plotenvelope}, which by default will produce a residual vs fits plot and a QQ-plot as discussed in this paper, with accompanying 95\% global envelopes constructed via calls to
{\verb+GET::global_envelope_test+}. The user can control the method of envelope construction and the number of sets of simulated residuals used. Default settings construct Studentized MAD global envelopes (\texttt{type="st"}) from $B=199$ sets of residuals. A scale-location plot can be constructed using the argument \texttt{which=3} (analogous to behaviour of \texttt{plot.lm} in base \texttt{R}).

As an example of how to use \texttt{plotenvelope}, \citet{Brooks2017} used generalised linear mixed models to study counts of salamander abundance across 23 sites, with four replicates at each site, in order to look for an impact of mining. Below we fit a Poisson log-linear model to the data with a random intercept for site (using the \texttt{glmmTMB} package), then check assumptions using all three of the proposed plots:
\begin{verbatim}
  library(glmmTMB)
  data(Salamanders)
  m1 <- glmmTMB(count ~ mined + (1|site),family=poisson, data=Salamanders)
  library(ecostats)
  plotenvelope(m1,which=1:3)
\end{verbatim}
We see for this dataset (Figure~\ref{fig:egPlots}) signs of overdispersion relative to the Poisson distribution, with points lying above their simulation envelope in the quantile plot (Figure~\ref{fig:egPlots}b), and an increasing trend that exceeds its bounds in the scale-location plot (Figure~\ref{fig:egPlots}c), for large fitted values.

The \texttt{ecostats::plotenvelope} function can in principle be applied to \emph{any} model object that has methods for the \texttt{simulate}, \texttt{residuals} and \texttt{predict} functions. To date it has been tested on common tools for fitting linear or generalised linear models, mixed models \citep{Bates2015,Brooks2017}, generalised additive models \citep{Wood2011}, and software for multivariate responses \citep{mvabundPaper}.

The method used to construct residuals and linear predictors in \texttt{plotenvelope} can be controlled by the user via the \texttt{resFunction} and \texttt{predFunction} arguments, which must be functions that take a fitted model as input and are expected to return residuals and linear predictors, respectively. These arguments default to the generic \texttt{residuals} and \texttt{predict} functions, respectively, but if unavailable for a particular model object then \texttt{resFunction} and \texttt{predFunction} could be used to define suitable alternatives. If desired, these functions could also be used to construct other types of plots beyond residual vs fitted value plots, by setting these functions such that they return values beyond residuals and/or fitted values. For example, we could use this function to obtain global envelopes around the expected trend on an added variable plot by setting \texttt{predFunction} to return the predictor of interest.

\begin{figure}
  \centering
  \includegraphics[width=\textwidth]{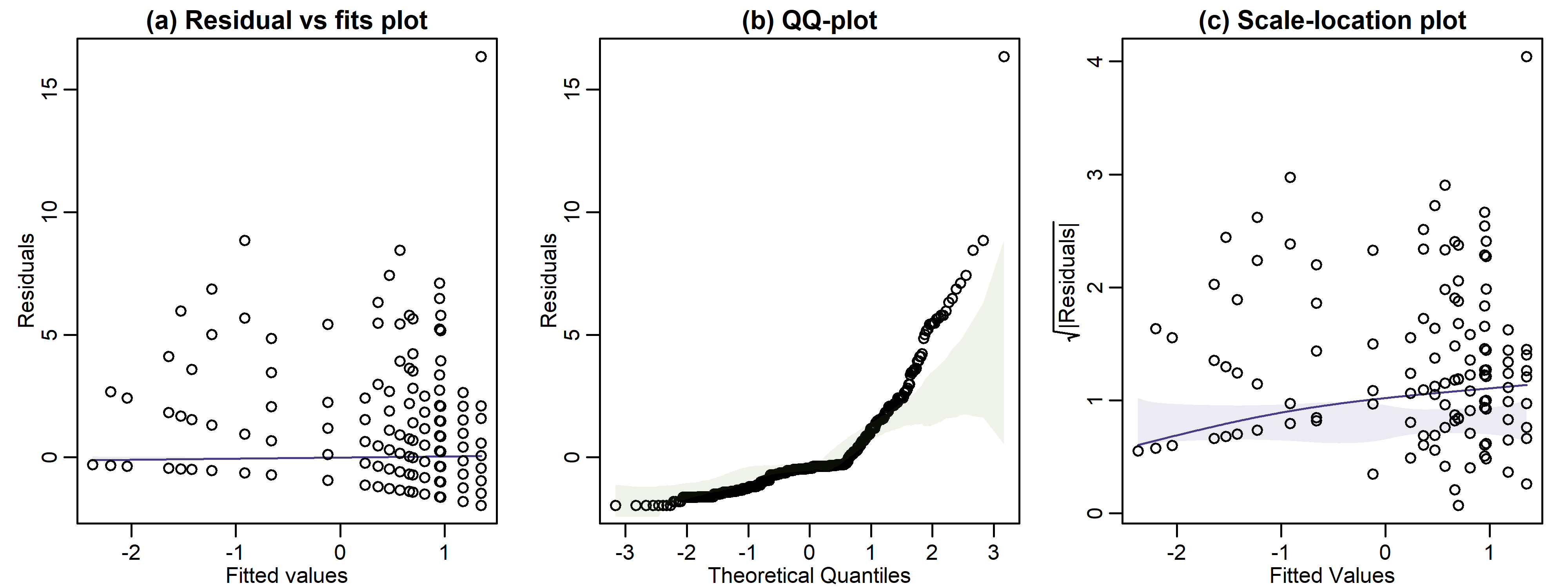}
  \caption{Example diagnostic plots with global simulation envelopes, constructed using \texttt{ecostats::plotenvelope}, for the \texttt{Salamanders} data \citep{Price2016} available in the \texttt{glmmTMB} package on \texttt{R} \citep{Brooks2017}. Three types of residual plot are presented: (a) a residual vs fitted values plot, (b) a QQ-plot, (c) a scale-location plot. A Poisson loglinear model with a random intercept was fitted using \texttt{glmmTMB}, and Pearson residuals computed from the fit. Global envelopes are the shaded areas, not that in (a) the envelope is so narrow that it is hardly visible.
  Note from (b) that many residuals on the QQ-plot fall outside their simulation envelope (taking unusually large values), and there is an increasing trend on the scale-location plot, which strays outside its envelope. Both of these trends are suggestive of overdispersion.}\label{fig:egPlots}
\end{figure}


\section{Simulations}

Although global simulation envelopes are proposed here primarily as a graphical aid, they can also be used to construct formal tests of assumptions. Simulations have been undertaken to study the performance of such tests, as compared to some other commonly available tests of model fit, for situations where we have $n$ univariate responses $y_i$, $i=1,\ldots,n$, assumed to be related to a single predictor $x_i$ via a linear function $\eta_i=\beta_0+x_i\beta_1$, and we fit one of three different types of model for the response $y_i$:
\begin{itemize}
\item[(a)] A linear model, $y_i\sim \mathcal{N}(\eta_i,\sigma^2)$, fitted on \texttt{R} using the \texttt{lm} function.
\item[(b)] A Poisson log-linear model, $y_i\sim \mathcal{P}(\exp\{\eta_i\})$, fitted on \texttt{R} using the \texttt{glm} function.
\item[(c)] A Poisson log-linear model with random intercept term, $y_{i}\sim \mathcal{P}(\exp\{\eta_i+\epsilon_{i \text{ mod } 5}\})$ where $\epsilon_j\sim\mathcal{N}(0,\omega^2)$ for $j\in \{0,1,2,3,4\}$, fitted on \texttt{R} using the \texttt{glmmTMB::glmmTMB} function. The normally distributed random intercept that has been added to the linear predictor can take five different values, which we cycle through repeatedly as we move through the dataset.
\end{itemize}
The \texttt{glmmTMB} package \citep{glmmTMBpackage}, used to fit Model (c), is an alternative to the widely used \texttt{lme4} package \citep{lme4Package}. It is used here because it tends to provide faster, more stable fits (while also having a broader range of possible variance structures and \texttt{family} arguments). This extra functionality is made available via Template Model Builder \citep{TMBpackage}, an environment for statistical model-building that makes use of automatic differentiation and \texttt{C++} for computational expediency.

Standardized residuals were used to evaluate fits of Model (a) via the \texttt{rstandard} function, which are normally distributed when assumptions are satisfied. For other models, we used deviance residuals \citep{Garciaetal12}, the default residual output for \texttt{glm}. In the mixed model setting (c), fitted values were estimated marginally with respect to the random effect. 
Note that in both (b) and (c), residuals are not normally distributed when assumptions are satisfied, motivating the use of simulation to construct global envelopes for residual plots.

Models (a-c) were each fitted to data simulated under settings similar to what was assumed by the corresponding model, but under three different configurations, corresponding to three different types of assumption violations. Specifically, in each case the linear predictor used in simulations had the form
\[ \eta_i = \beta_0 + 4 x_i + \beta_2 x_i^2, \]
but data were simulated under the following three different configurations:
\begin{description}
\item[Assumptions satisfied] Data were simulated from Models (a-c) as described previously, with $\beta_0=-2$, $\beta_2=0$ and $\sigma=0.25$ (and for Model (c), $\omega=1$).
\item[Mixture distribution]
A 9:1 mixture of responses was simulated as follows:
\begin{itemize}
\item[(a)] $y_i\sim 0.9 \mathcal{N}(\eta_i,\sigma^2)+0.1\mathcal{N}(\eta_i,(4\sigma)^2)$
\item[(b)] $y_i\sim 0.9 \mathcal{P}(\exp\{\eta_i\})+0.1\mathcal{P}(4\exp\{\eta_i\})$
\item[(c)] $y_i\sim 0.9 \mathcal{P}(\exp\{\eta_i+\epsilon_{i \text{ mod } 5}\})+0.1\mathcal{P}(4\exp\{\eta_i+\epsilon_{i \text{ mod } 5}\})$
\end{itemize}
with $\beta_0=-2$, $\beta_2=0$, $\sigma=0.25$ and $\omega=1$ as before. Note that each of Models (a-c) correctly specify the functional form of their respective model for the marginal mean -- linearity is preserved in (a), and log-linearity in (b-c) -- but distributional assumptions of each model have been violated.
\item[Quadratic response] Data were simulated from Models (a-c) with $\beta_0=1$ and $\beta_2=-4$ (and, as previously, $\sigma=0.25$ and $\omega=1$). This produced (on the linear predictor scale) a parabolic response whose axis of symmetry was at the centre of the $x_i$, and whose values for the $\eta_i$ were centred on similar values to those produced by other simulation scenarios.
\end{description}

This gives a total of nine simulation scenarios, consisting of three models (a-c) fitted under each of three assumption violation configurations -- when assumptions are satisfied, when distributional assumptions are violated, and when the mean model is not correctly specified. Under each scenario, 1000 datasets were simulated at each of four different sample sizes -- $n\in\{10,20,40,80\}$. The following tests were compared, in terms of their rejection rate at the $\alpha=0.05$ level:
\begin{itemize}
\item Global simulation envelopes around smoothers on a residual vs fits plot or scale-location plot as described in the previous section (using Studentized MAD envelopes). We expected the residual vs fits plot to be useful for detecting assumption violations in \textbf{Quadratic response} simulations, and the scale-location plot to be more useful for \textbf{Mixture distribution} simulations.
\item Global simulation envelopes around a normal quantile (QQ-) plot or a probability (PP-) plot (using Studentized MAD envelopes). We expected these to be most useful in \textbf{Mixture distribution} simulations.
\item For Model (a), a Shapiro-Wilks test was computed using the \texttt{shapiro.test} function on \texttt{R}.
\item For Models (b-c), the maximized log-likelihood was used as a goodness-of-fit test statistic, and compared to its null distribution as estimated from simulated datasets $\mathbf{y}_b$ described previously. Lack-of-fit is indicated if the observed value $\log\mathcal{L}(\boldsymbol{\theta},\mathbf{y})$ is small compared to values computed for data simulated under the assumed model, $\log\mathcal{L}(\boldsymbol{\theta},\mathbf{y}_b)$.
\end{itemize}
In all cases, simulation envelopes were constructed using 999 simulated datasets (in addition to the observed data), and interpreted as a hypothesis test by asking whether the observed data (or smoother) fell outside its 95\% global envelope at any stage. Because this whole process was repeated 1000 times, for each sample size and simulation setting, simulations were quite computationally intensive and were undertaken on a computational cluster \citep{Katana}, requiring several thousand hours in total computation time on processors with 20-60GB RAM.

Results for these nine simulation scenarios are summarized in Figure~\ref{fig:simRes}. Note firstly that the global envelopes all adequately control Type I error (Figure~\ref{fig:simRes}, left), although some methods were conservative. \citet{Myllymaki2017} also saw conservativeness in their simulations, and noted that it is often observed empirically in Monte Carlo testing of composite hypotheses.

When looking at power of different methods to detect assumption violations, the most striking result was that the most effective tools were simulation envelopes around a QQ-plot or a residual vs fits plot, for mixture distributions (Figure~\ref{fig:simRes}, centre) and quadratic response (Figure~\ref{fig:simRes}, right), respectively. Note that global envelopes on a QQ-plot performed comparably to a Shapiro-Wilks test at finding violations of normality (Figure~\ref{fig:simRes}a, centre) and better than a deviance test at detecting violations of a Poisson assumption (Figure~\ref{fig:simRes}b-c, centre). When there was a quadratic response to predictors, a global envelope around a smoother on a residual vs fits plot was by far the most effective method (Figure~\ref{fig:simRes}, right) for all three models.

Note that for non-Gaussian models (Figure~\ref{fig:simRes}b-c) there was some interaction between non-normality and non-linearity, with residual vs fits plots having some ability to detect mixture distributions and a QQ-plot having some power under non-linearity. Mean and variance parameters are orthogonal in the linear model, hence it is not surprising that diagnostics designed to target the mean trend performed poorly when distributional assumptions were violated, and diagnostics designed to target distributional assumptions performed poorly when the mean model was incorrect. In the Poisson model, in contrast, a missing quadratic term introduces overdispersion, which can be detected on a QQ-plot (Figure~\ref{fig:simRes}b-c, right).

\begin{figure}
  \centering
  \includegraphics[width=\textwidth]{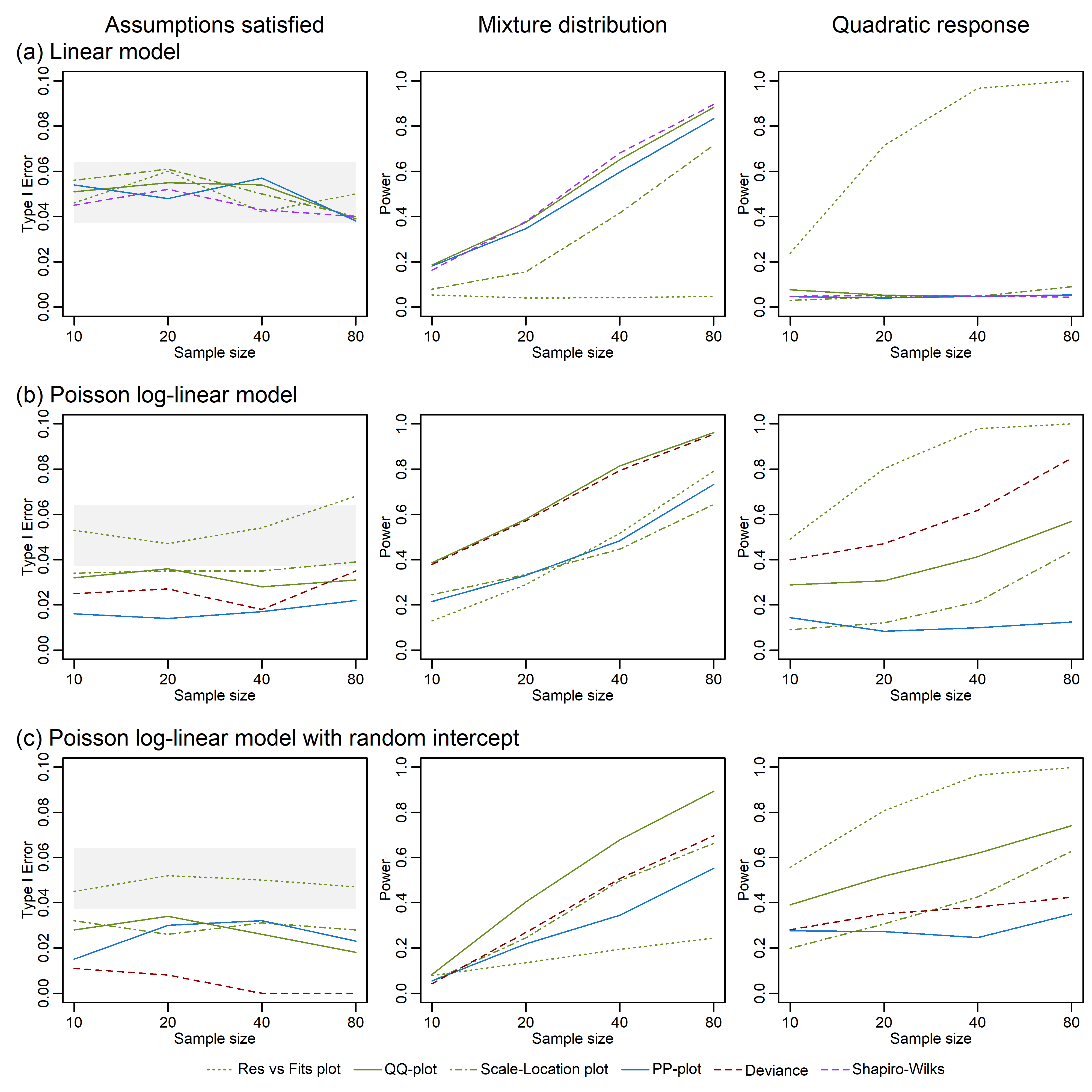}
  \caption{Simulation results when fitting Models (a-c) to data simulated under the assumed model (left column), from a mixture distribution (centre) or with a quadratic response to the predictor (right). In (a), the grey area is a 95\% global envelope around the observed Type I error of an exact test, at significance level of 0.05, in this simulation design. Note in particular that all global simulation envelopes adequately control Type I error (left), that envelopes around QQ-plots (green solid line) were most effective at finding violations of distributional assumptions (center), and envelopes around a smoother on a residual vs fits plot (green dotted line) was most effective at finding violations of linearity (right).}\label{fig:simRes}
\end{figure}

As a secondary result, note that the scale on which diagnostic tools are plotted was important -- QQ-plots consistently outperformed PP-plots at detecting assumption violations. This was expected since assumption violations occurred in the tails of the distribution, and plotting residuals on the quantile scale puts the emphasis on the tails, whereas a PP-plot transforms residuals to fit in the unit interval. 

A global envelope around the smoother on a scale-location plot was the least effective of the proposed methods. As expected, it was useful for identifying distributional assumption violations (Figure~\ref{fig:simRes}, centre), but was noticeably less effective than a QQ-plot, or a Shapiro-Wilks test of normality (Figure~\ref{fig:simRes}a, centre), or a deviance-based test for Poisson data (Figure~\ref{fig:simRes}b, centre).

\section{Discussion}
In this paper we showed how recent advances in tools for global simulation envelopes can be used to aid in the interpretation of common residual plots. Previously, global simulation envelopes had been proposed for QQ-plots \citep{Rosenkrantz2000,Aldor-Noiman2013}, assuming residuals were Gaussian and independently and identically distributed (``iid"). However, one or both of the Gaussian and iid assumptions will not be satisfied for residuals from many types of regression model, an issue which we resolve here by using simulation to explore the expected distribution of residuals and residual functions, by simulating data from the true model and recomputing residuals. Software is provided (the \texttt{plotenvelope} function in the \texttt{ecostats} package) to compute global simulation envelopes for various types of residual plot for a broad class of regression models, and illustrated its effectiveness by simulation for a few common types of regression model.

The intention is that global envelopes be used as a visual tool to help the analyst determine whether any departures from expected trends are large relative to what would be expected if assumptions were true. This could be especially useful as a teaching tool, for students not experienced with interpreting residual plots -- in fact, the motivation for this work arose for the author while writing a text for ecologists on data analysis \citep{WartonBook}. While not the primary intention, the global envelope approach can also be interpreted as a formal test of assumptions, which was helpful as a way to study the effectiveness of the tool. It was interesting to see in simulations that the proposed methods tended to compare favorably to common goodness-of-fit tests, suggesting that using global envelopes on residual plots is relatively effective as a way to detect violations of assumptions.

Simulations reinforce the idea that different diagnostic plots detect different types of assumption violations, by design, and as such it is a good idea to use multiple diagnostic plots when interrogating multiple assumptions. Beyond the plots discussed here, one could also look at response to predictors, as advocated for by \citet{CookMarginal}. The \texttt{plotenvelope} function could be used to produce plots against a predictor of particular interest by setting \texttt{predFunction} to be a function that returned the predictor in question.

The performance of the global envelope tests depend heavily on the type of residual used to construct plots, the scale on which residuals are plotted, and if envelopes are constructed around smoothers, the method of smoother construction, as discussed in more detail below.

The way residuals are constructed is clearly an important consideration when constructing any residual plot. In preliminary investigations, we found for example that when fitting a generalized linear mixed model, residual plots had difficulty detecting violations of distributional assumptions (Figure~\ref{fig:simRes}c, centre) when using unscaled residuals ($y_i-\widehat{\mu}_i$) rather than deviance residuals. A potentially useful alternative way to construct residuals for non-Gaussian models is randomized quantile residuals \citep{DunnSmyth}, which introduce jittering to discrete data to ensure that residuals computed from the true model will be marginally standard normal, for any parametric model. Such residuals can be computed on \texttt{R} for generalized linear mixed models using the \texttt{statmod::qres} function, and can be computed by simulation from broader classes of distribution using the DHARMa package \citep{DHARMa}.

The scale on which residuals are plotted can also be important to performance of global envelopes, as evidenced by the differences in performance between QQ-plots and PP-plots (Figure~\ref{fig:simRes}, centre and right). While PP-plots are often used as a diagnostic tool \citep[for example]{DHARMa}, they were consistently outperformed by QQ-plots
. These plots differ only in their scale, with PP-plots mapping residuals to the unit interval (via $\Phi^{-1}(e_i)$), then comparing to standard uniform rather than to standard normal quantiles. But violations of assumptions typically are expressed in the tails of a distribution, with the occasional very large (in magnitude) residual. When converted to the unit interval, such outliers become values that are unusually close to zero or one, which does not provide a strong visual cue. For example, in Figure~\ref{fig:egPlots}, a striking feature is the outlying Pearson residual, with a value of about 15, which had an expected normal quantile closer to 3. On a PP-plot, this would be expressed as a probability that was equal to one up to machine error, as compared to an expected probability of 0.998. Mapping residuals to the unit interval compresses the regions of primary interest to narrow ranges of values, making it difficult to diagnose violations of assumptions. For this reason \citet{DunnSmyth} advocated mapping quantile residuals to the normal distribution.

Preliminary investigations also suggested that when constructing a global envelope around a smoother, performance of envelope tests was sensitive to the method of smoother construction. When using the default \texttt{mgcv::gam} method, which minimizes a generalized cross validation (GCV) measure, our smoothers had quite poor performance at detecting assumption violations. The relatively good performance seen in Figure~\ref{fig:simRes} was achieved through use of maximum likelihood estimation to fit smoothers (\texttt{method="ml"}), which is noticeably more computationally intensive, but known to perform better at estimating smooth functions, largely eliminating the chance of ``severe undersmoothing failures" seen in GCV techniques \citep{Wood2011}.

Finally, it is worth reminding the practitioner that significant violations of assumptions are not necessarily of practical importance. Linear models for example are known to be quite robust to violations of normality due to the Central Limit Theorem, especially at large sample sizes \citep[see][for example]{Lumley2002}. Large sample sizes however lead to narrower simulation envelopes and a greater chance of detecting assumption violations, in the very situation where they are less important. It should always be kept in mind by the practitioner that global envelopes are graphical tools that show how large deviations are relative to those that might be expected due to sampling error, but they do not indicate if deviations are large enough to be of practical importance. Effect size is best measured by looking at the magnitude of deviations from expected behaviour \citep[for example]{Kim2018}, rather than just looking at whether these deviations stray outside a global envelope.

\subsection*{Disclosure statement}
The author reports there are no competing interests to declare.

\subsection*{Acknowledgements}
The author's work was supported by an Australian Research Council Discovery Projects scheme (DP180103543). Thanks to anonymous reviewers for suggestions that improved this manuscript.


\begin{thebibliography}{24}
\newcommand{\enquote}[1]{``#1''}
\expandafter\ifx\csname natexlab\endcsname\relax\def\natexlab#1{#1}\fi
\expandafter\ifx\csname url\endcsname\relax
  \def\url#1{{\tt #1}}\fi
\expandafter\ifx\csname urlprefix\endcsname\relax\def\urlprefix{URL }\fi

\bibitem[{Aldor-Noiman et~al.(2013)Aldor-Noiman, Brown, Buja, Rolke, and
  Stine}]{Aldor-Noiman2013}
Aldor-Noiman, S., Brown, L.~D., Buja, A., Rolke, W., and Stine, R.~A.
\newblock \enquote{The Power to See: A New Graphical Test of Normality
  Statistical Computing and Graphics The Power to See: A New Graphical Test of
  Normality.} (2013).

\bibitem[{Bates et~al.(2015{\natexlab{a}})Bates, Mächler, Bolker, and
  Walker}]{Bates2015}
Bates, D., Mächler, M., Bolker, B.~M., and Walker, S.~C.
\newblock \enquote{Fitting Linear Mixed-Effects Models Using lme4.}
\newblock {\em Journal of Statistical Software\/}, 67:1--48
  (2015{\natexlab{a}}).

\bibitem[{Bates et~al.(2015{\natexlab{b}})Bates, Mächler, Bolker, and
  Walker}]{lme4Package}
---.
\newblock \enquote{Fitting Linear Mixed-Effects Models Using lme4.}
\newblock {\em Journal of Statistical Software\/}, 67:1--48
  (2015{\natexlab{b}}).

\bibitem[{Brooks et~al.(2017{\natexlab{a}})Brooks, Kristensen, van Benthem,
  Magnusson, Berg, Nielsen, Skaug, Mächler, and Bolker}]{Brooks2017}
Brooks, M.~E., Kristensen, K., van Benthem, K.~J., Magnusson, A., Berg, C.~W.,
  Nielsen, A., Skaug, H.~J., Mächler, M., and Bolker, B.~M.
\newblock \enquote{glmmTMB balances speed and flexibility among packages for
  zero-inflated generalized linear mixed modeling.}
\newblock {\em R Journal\/}, 9:378--400 (2017{\natexlab{a}}).

\bibitem[{Brooks et~al.(2017{\natexlab{b}})Brooks, Kristensen, van Benthem,
  Magnusson, Berg, Nielsen, Skaug, Mächler, and Bolker}]{glmmTMBpackage}
---.
\newblock \enquote{glmmTMB balances speed and flexibility among packages for
  zero-inflated generalized linear mixed modeling.}
\newblock {\em R Journal\/}, 9:378--400 (2017{\natexlab{b}}).

\bibitem[{Cook and Weisberg(1997)}]{CookMarginal}
Cook, R.~D. and Weisberg, S.
\newblock \enquote{Graphics for Assessing the Adequacy of Regression Models.}
\newblock {\em Journal of the American Statistical Association\/}, 92:490--499
  (1997).

\bibitem[{Davison and Hinkley(1997)}]{davison1997bootstrap}
Davison, A.~C. and Hinkley, D.~V.
\newblock {\em Bootstrap methods and their application\/}.
\newblock 1. Cambridge university press (1997).

\bibitem[{Dunn and Smyth(1996)}]{DunnSmyth}
Dunn, P.~K. and Smyth, G.~K.
\newblock \enquote{Randomized Quantile Residuals.}
\newblock {\em Journal of Computational and Graphical Statistics\/}, 5:236--244
  (1996).

\bibitem[{García et~al.(2012)García, Víctor, Yohai, Ben, and
  Yohai}]{Garciaetal12}
García, M., Víctor, B.~., Yohai, J., Ben, M.~G., and Yohai, V.~J.
\newblock \enquote{Quantile–Quantile Plot for Deviance Residuals in the
  Generalized Linear Model.}
\newblock {\em Journal of Computational and Graphical Statistics\/}, 13:36--47
  (2012).

\bibitem[{Hartig(2022)}]{DHARMa}
Hartig, F.
\newblock {\em DHARMa: Residual Diagnostics for Hierarchical (Multi-Level /
  Mixed) Regression Models\/} (2022).
\newblock R package version 0.4.5.
\newline\urlprefix\url{https://CRAN.R-project.org/package=DHARMa}

\bibitem[{Kim and Cribbie(2018)}]{Kim2018}
Kim, Y.~J. and Cribbie, R.~A.
\newblock \enquote{ANOVA and the variance homogeneity assumption: Exploring a
  better gatekeeper.}
\newblock {\em British Journal of Mathematical and Statistical Psychology\/},
  71:1--12 (2018).

\bibitem[{Kristensen et~al.(2016)Kristensen, Nielsen, Berg, Skaug, and
  Bell}]{TMBpackage}
Kristensen, K., Nielsen, A., Berg, C.~W., Skaug, H., and Bell, B.~M.
\newblock \enquote{TMB: Automatic Differentiation and Laplace Approximation.}
\newblock {\em Journal of Statistical Software\/}, 70:1--21 (2016).

\bibitem[{Lumley et~al.(2002)Lumley, Diehr, Emerson, and Chen}]{Lumley2002}
Lumley, T., Diehr, P., Emerson, S., and Chen, L.
\newblock \enquote{The Importance of the Normality Assumption in Large Public
  Health Data Sets.}
\newblock {\em Annual Review of Public Health\/}, 23:151--169 (2002).

\bibitem[{Moral et~al.(2017)Moral, Hinde, and Demétrio}]{Moral2017}
Moral, R.~A., Hinde, J., and Demétrio, C.~G.
\newblock \enquote{Half-normal plots and overdispersed models in R: The hnp
  package.}
\newblock {\em Journal of Statistical Software\/}, 81 (2017).

\bibitem[{Myllymäki et~al.(2017)Myllymäki, Mrkvička, Grabarnik, Seijo, and
  Hahn}]{Myllymaki2017}
Myllymäki, M., Mrkvička, T., Grabarnik, P., Seijo, H., and Hahn, U.
\newblock \enquote{Global envelope tests for spatial processes.}
\newblock {\em Journal of the Royal Statistical Society: Series B (Statistical
  Methodology)\/}, 79:381--404 (2017).

\bibitem[{Price et~al.(2016)Price, Muncy, Bonner, Drayer, and
  Barton}]{Price2016}
Price, S.~J., Muncy, B.~L., Bonner, S.~J., Drayer, A.~N., and Barton, C.~D.
\newblock \enquote{Effects of mountaintop removal mining and valley filling on
  the occupancy and abundance of stream salamanders.}
\newblock {\em Journal of Applied Ecology\/}, 53:459--468 (2016).

\bibitem[{{R Core Team}(2022)}]{CRAN}
{R Core Team}.
\newblock {\em R: A Language and Environment for Statistical Computing\/}.
\newblock R Foundation for Statistical Computing, Vienna, Austria (2022).
\newline\urlprefix\url{https://www.R-project.org/}

\bibitem[{Ripley(1976)}]{Ripley1976}
Ripley, B.~D.
\newblock \enquote{The second-order analysis of stationary point processes.}
\newblock {\em Journal of Applied Probability\/}, 13:255--266 (1976).

\bibitem[{Rosenkrantz(2000)}]{Rosenkrantz2000}
Rosenkrantz, W.~A.
\newblock \enquote{Confidence Bands for Quantile Functions: A Parametric and
  Graphic Alternative for Testing Goodness of Fit.}
\newblock {\em American Statistician\/}, 54:185--190 (2000).

\bibitem[{{UNSW Sydney}(2022)}]{Katana}
{UNSW Sydney}.
\newblock {\em Katana\/}.
\newblock PVC (Research Infrastructure), UNSW Sydney (2022).
\newline\urlprefix\url{https://doi.org/10.26190/669x-a286}

\bibitem[{Wang et~al.(2012)Wang, Naumann, Wright, and Warton}]{mvabundPaper}
Wang, Y., Naumann, U., Wright, S.~T., and Warton, D.~I.
\newblock \enquote{\texttt{mvabund} -- an \texttt{R} package for model-based
  analysis of multivariate abundance data.}
\newblock {\em Methods in Ecology and Evolution\/}, 3:471--474 (2012).

\bibitem[{Warton(2022)}]{WartonBook}
Warton, D.~I.
\newblock {\em Eco-Stats: Data Analysis in Ecology, from $t$-tests to
  multivariate abundances\/}.
\newblock Springer Nature (2022).

\bibitem[{Wood(2003)}]{Wood2003}
Wood, S.~N.
\newblock \enquote{Thin plate regression splines.}
\newblock {\em Journal of the Royal Statistical Society: Series B (Statistical
  Methodology)\/}, 65:95--114 (2003).

\bibitem[{Wood(2011)}]{Wood2011}
---.
\newblock \enquote{Fast stable restricted maximum likelihood and marginal
  likelihood estimation of semiparametric generalized linear models.}
\newblock {\em Journal of the Royal Statistical Society: Series B (Statistical
  Methodology)\/}, 73:3--36 (2011).

\end{thebibliography}
\end{document}